\documentclass[12pt]{article}
\usepackage{amsfonts}
\usepackage{amsmath}
\usepackage{amssymb}
\usepackage{graphicx}
\usepackage{epsfig}

\begin{document}

\begin{center}
{ \large \bf A crucial hypothesis for Inflation}
\end{center}


\vspace{12pt}

\begin{center}
{\large {\em Giandomenico Palumbo}}
\end{center}

\vspace{6pt}

\begin{center}
Dipartimento di Fisica Nucleare e Teorica,
Universit\`a degli Studi di Pavia and
Istituto Nazionale di Fisica Nucleare, Sezione di Pavia\\
via A. Bassi 6, 27100 Pavia (Italy)\\
E-mail: giandomenico.palumbo@pv.infn.it
\end{center}

\vspace{6pt}

\begin{center}
This essay has received an Honorable Mention from the Gravity
Research Foundation-2011 Awards for Essays on Gravitation.
\end{center}

\vspace{6pt}

\vspace{6pt}

\begin{abstract}
 As is well known there are many different inflationary models that can explain
 the accelerated expansion occurred in the early Universe.
 It is possible that there exists a fundamental property of this period
  that could provide the corresponding field theory.
 Our hypothesis is that the Ricci scalar should assume
 a positive constant value during Inflation.
 Considering a single scalar field that drives the
 inflationary phase,
 we obtain uniquely the scalar potential and the coupling term
 between the scalar field and spacetime geometry.
 For this potential, that is
 well known in the inflationary scenario,
 the theoretical prediction of the spectral index $n_{s}$ is in perfect
agreement with the experimental data.
 Interestingly in two dimensions,
 our model is equivalent to Liouville
 gravity.
\end{abstract}

\vspace{4 cm}

\vspace{6pt}

\vfill
\newpage

\section{Introduction}

\vspace{0.5cm}

The aim of Cosmology is to explain the origin, dynamics and
evolution of the Universe at large scales. For large scales we
mean those for which galaxies and clusters of galaxies can be
considered as point-like objects. A cosmological model is a model
that attempts to explain the dynamics of these material points in
agreement with astronomical observations. The most powerful model
is the Cosmological Standard Model. Its basic assumption is that
at large scales and at all times of expansion of the Universe, the
matter (regarded as a perfect fluid) is spatially distributed in a
homogeneous and isotropic manner.

 At these scales, gravity is the dominant force and it is described
 by Einstein's General Relativity and
 the space-time metric is the Friedmann-Robertson-Walker (FRW) metric.
 The Cosmological Standard Model explains the abundance of light elements,
 the recession of galaxies, the thermodynamic state of the observable universe.
 However, it also has some inconsistencies as the problem of homogeneity,
 flatness, and that concerns the birth and formation of large scale structures.

 A. Guth realized \cite{Guth}, that considering in the early Universe a very short
 period during which the scale factor expands exponentially, it was possible
 to solve the three problems mentioned. This phase is called the inflationary
 phase.
 To obtain this expansion, we must go beyond the assumption of perfect fluid for matter.
 It is known that at high temperatures the classical description of
  matter as an ideal gas is not valid. We expect that a fair description of the early
 Universe can be determined using a valid theory of matter at high energies.
  One possibility is to use a scalar field. In this context an inflation model
   must explain the anisotropies of the CMB (the Cosmic Microwave Background radiation)
    and the reheating  \cite{Lyth}.
The reheating is the process by which inflation ends. Of course
the end of inflation is crucial because only in a period of the
Universe when the expansion is not accelerated we can have the
baryogenesis.

Thus over the years a myriad of inflationary theories have been
developed \cite{Lemoine}, with many different self-interaction
potentials and there is not yet a standard inflationary model.

The goal in our work is to start with a precise assumption on
Inflation, obtaining uniquely the relative scalar field action. We
consider a single scalar field that drives Inflation and the
crucial hypothesis is that in the inflationary period, the
spacetime lorentzian manifold has a positive constant Ricci
scalar. The demand for a positive constant Ricci scalar is not a
request so unusual.
 The inflationary phase is in fact also called de Sitter phase. In
 a pure de Sitter phase, a cosmological
  constant dominates the expansion dynamics and the spacetime is
  called de
  Sitter spacetime (vacuum solution of Einstein's field equations
  with a positive cosmological constant). In this spacetime
  the Ricci scalar is constant. We generalize this propriety about
   Ricci scalar to a
   spacetime where there is also a scalar field.

  In the next section, we
  derive uniquely
   the coupling term between the scalar field and
  the Ricci scalar and the inflationary potential.
   This scalar potential is the double-well potential and it is well known in the
  study of inflation \cite{Albrecht} \cite{Vilenkin}
\cite{Riotto} \cite{Linde} \cite{Boyano} \cite{Rehman}
\cite{Shapo}.
  In this model, the theoretical
prediction of the spectral index is in perfect agreement with its
experimental value \cite{Komatsu} \cite{Linde} \cite{Boyano}
\cite{Rehman}.

  Finally we show that
  in the limit  $D = 2$  the model is equivalent to Liouville gravity,
  an important model of two dimensional gravity that derives from
  the conformal anomaly of non-critical string theory \cite{Nakayama}.

\vspace{0.5cm}

\section{The Model}

\vspace{0.2cm}

Let's consider the action of a single scalar field in D
dimensions:

\begin{equation}
  S_{D}=\int d^{D}x \sqrt{-g}\left[\frac{1}{2}\partial^{\mu}\phi\partial_{\mu}\phi
  +f(\phi)R+U(\phi)\right]
\end{equation}

\vspace{0.2cm}

where $g_{\mu\nu}$ is a pseudo-riemannian metric, $f(\phi)R$ is
the coupling term between the scalar field and spacetime and
$U(\phi)$ is the scalar potential.

With this action, we derive the equations of motion:

\begin{equation}
  \square \phi-f'(\phi)R-U'(\phi)=0
\end{equation}

\begin{eqnarray}
  f(\phi)\left(R_{\mu\nu}-\frac{1}{2}g_{\mu\nu}R\right)+\frac{1}{2}
  \partial_{\mu}\phi\partial_{\nu}\phi-\frac{1}{4}g_{\mu\nu}
  \partial^{\gamma}\phi\partial_{\gamma}\phi-f'(\phi)\partial_{\mu}
  \partial_{\nu}\phi && \nonumber\\-f''(\phi) \partial_{\mu}\phi\partial_{\nu}\phi
  +g_{\mu\nu}f'(\phi)\square \phi+g_{\mu\nu}f''(\phi)\partial^{\gamma}\phi
  \partial_{\gamma}\phi-\frac{1}{2}g_{\mu\nu}U(\phi)=0
\end{eqnarray}

\vspace{0.2cm}

where $f'(\phi)=\frac{d}{d\phi}f(\phi)$ and
$U'(\phi)=\frac{d}{d\phi}U(\phi)$.

\vspace{0.2cm}

The first equation is the generalized Klein-Gordon equation. We
calculate the trace of the second equation and after some
manipulations, we obtain

\begin{eqnarray}
  \square\phi-\frac{\left(\frac{D-2}{2}\right)f(\phi)R+
  \left(\frac{D}{2}\right) U(\phi)}{(D-1)f'(\phi)}+
  \frac{\left[(D-1)f''(\phi)-\frac{D-2}{4}\right]
  \partial^{\gamma}\phi\partial_{\gamma}\phi}{(D-1)f'(\phi)}
  =0
\end{eqnarray}

\vspace{0.2cm}

Our model is based on the hypothesis that during the inflationary
period, the Ricci scalar $R$ assumes a positive constant value:

\begin{equation}
  R=\alpha^{2}
\end{equation}

\vspace{0.2cm}

then we consider only a subclass of metrics of $g_{\mu\nu}$,
solution of equations (2) and (3), that admits this condition.

Supposing to know explicitly the metric and replacing $R$ with
$\alpha^{2}$ in (2) and (4) thanks to the relation (5), we have a
system of two partial differential equations for the same field
$\phi$. A simple way, for being sure that the system admit a
solution, is to consider the equations (2) and (4) as the same
equation. We have the following conditions:

\begin{equation}
(D-1)f''(\phi)-\frac{D-2}{4}=0
\end{equation}

\begin{equation}
f'(\phi)\alpha^{2}+U'(\phi)=\frac{\left(\frac{D-2}{2}\right)f(\phi)\alpha^{2}+
  \left(\frac{D}{2}\right) U(\phi)}{(D-1)f'(\phi)}
\end{equation}

\vspace{0.2cm}

 and the relative solutions are respectively:

\begin{eqnarray}
  f(\phi)=\frac{1}{4}\left(\frac{D-2}{D-1}\right)\frac{\phi^{2}}{2}
  +\beta\phi+\delta
\end{eqnarray}

\begin{eqnarray}
  U(\phi)=C\left[\left(\frac{D-2}{2D}\right)\phi
  +2\left(\frac{D-1}{D}\right)\beta\right]^{\frac{2D}{D-2}}+&\nonumber
  \\\alpha^{2}\left[2\left(\frac{D-1}{D}\right)\beta^{2}
  -\left(\frac{D-2}{D}\right)\delta\right]
\end{eqnarray}

\vspace{0.2cm}

where $\beta$, $\delta$ and C are generic constants and
$\xi=\frac{1}{4}\left(\frac{D-2}{D-1}\right)$ is the conformal
coupling in D dimensions.

Placing $\beta=\frac{(4-D)(3-D)}{2}\beta'$ (where $\beta'$ is
another constant), we get the following identifications:

\begin{eqnarray}
  \delta=-\frac{1}{16 \pi G}\hspace{1.0cm}
  \alpha^{2}=\frac{2D}{D-2} \Lambda
\end{eqnarray}

\vspace{0.2cm}

where $G$ is the Newton's constant and $\Lambda$ is a cosmological
constant ($\Lambda$ has not the same numerical value of the
cosmological constant of $\Lambda$CDM model). The action (1) for
$D=4$ is

\begin{eqnarray}
  S_{4}=\int d^{4}x \sqrt{-g}\left[-\frac{1}{16\pi G}\left(R-2\Lambda\right)
  +\frac{1}{2}\partial^{\mu}\phi\partial_{\mu}\phi
  +\frac{1}{12}\phi^{2}R
  -C'\phi^{4}\right]
\end{eqnarray}

\vspace{0.2cm}

where $C'=-\frac{C}{4^{4}}$

Thus we have determined exactly the total action for our scalar
field.

Moreover the coupling term $\phi^{2}R$ and the Einstein-Hilbert
term represent two potential terms because the Ricci scalar is
constant and we can replace it with $\alpha^{2}$. Thus in the
action $S_{4}$ there is a kinetic term and a total potential
$V(\phi)$.

The Lagrangian density is
$\mathcal{L}=\sqrt{-g}\left[\frac{1}{2}\partial^{\mu}\phi\partial_{\mu}\phi
-V(\phi)\right]$  with

\begin{eqnarray}
  V(\phi)=\frac{\Lambda}{8 \pi G}
  -\frac{\Lambda}{3}\phi^{2}+C'\phi^{4}
\end{eqnarray}

\vspace{0.2cm}

where we have used the relation (10) between $\alpha^{2}$ and
$\Lambda$.

In the context of Cosmological Standard Model, we consider the FRW
flat metric: \hspace{0.2cm} $ds^{2}=-dt^{2}+a(t)^{2}d\vec{x}^{2}$,
\hspace{0.1cm}   where $a(t)$ is the scale factor.

With this metric, the equations (5) and (2) become

\begin{eqnarray}
  R=6\left[\frac{\ddot{a}}{a}+\left(\frac{\dot{a}}{a}\right)^{2}\right]=
  6(2H^{2}+\dot{H})=\alpha^{2}
\hspace{1.0cm}\ddot{\phi}+3H\dot{\phi}+\frac{d V(\phi)}{d\phi}=0
\end{eqnarray}

\vspace{0.2cm}

where $H=\frac{\dot{a}}{a}$

A solution of the first relation gives us the scale factor $a(t)$
that represents the accelerated expansion:

\begin{eqnarray}
  a(t)=a_{0}\sqrt{\cosh \sqrt{\frac{4\Lambda}{3}}\hspace{0.1cm}t }
\end{eqnarray}

\vspace{0.2cm}

where $a_{0}=a(0)$.

Moreover if we introduce the (non standard) Planck energy scale
$M_{P}=\sqrt{\frac{3}{4 \pi G}}$ ($c=\hbar=1$) and the energy
scale of Inflation $M_{I}=\alpha=\sqrt{4\Lambda}$ and considering
$C'=\frac{1}{4!}\left(\frac{M_{I}}{M_{P}}\right)^{2}$, we can
rewrite the potential (12) as

\begin{eqnarray}
V(\phi)=\frac{1}{4!}\left(M_{I}M_{P}\right)^{2}\left[1-\left(
\frac{\phi}{M_{P}}\right)^{2}\right]^{2}
\end{eqnarray}

\vspace{0.2cm}

and the hierarchy condition $M_{I}\ll M_{P}$ \cite{Vega}
guarantees that $C'\ll 1$.

This scalar potential is well known in the inflationary scenario.
It is the double-well (Landau-Ginzburg) potential and it has been
studied in different approaches \cite{Albrecht} \cite{Vilenkin}
\cite{Riotto} \cite{Linde} \cite{Boyano} \cite{Rehman}
\cite{Shapo}.

In this model, the fundamental result in the context of slow-roll
Inflation \cite{Lyth}, is that the theoretical prediction of the
spectral index $n_{s}$ \cite{Linde} \cite{Boyano} \cite{Rehman} is
in perfect agreement with its experimental value \cite{Komatsu}.

Our final consideration concerns the fact that in (9) for $D=2$
and fixing $\beta=1$, there is the finite limit

\begin{eqnarray}
  \lim_{D\rightarrow2}\left[\left(\frac{D-2}{2D}\right)\phi
  +\frac{(D-1)(4-D)(3-D)}{D}
  \right]^{\frac{2D}{D-2}}=e^{\phi}
\end{eqnarray}

\vspace{0.2cm}

and then in two dimensions, the total action is

\begin{eqnarray}
  S_{2}=\int d^{2}x \sqrt{-g}\left[\frac{1}{2}\partial^{\mu}\phi\partial_{\mu}\phi
  +\phi R+C e^{\phi}+\alpha^{2}\right]
\end{eqnarray}

\vspace{0.2cm}

It coincides with the action of Liouville gravity that is an
important action in the context of non-critical string theory and
it is also a relevant model of gravity in two dimensions
\cite{Nakayama}. This link with the Liouville gravity could be an
first indication of a deeper connection between our
phenomenological theory and a more fundamental theory.

In conclusion, starting by a simple hypothesis about spacetime
geometry during the inflationary period, we have obtained the
precise action of scalar field with a specific potential, where
the theoretical prediction of the spectral index $n_{s}$ is in
perfect agreement with the experimental data.



\vspace{13.5cm}

\begin{thebibliography}{99}

\bibitem{Guth}
A. H. Guth, \textit{Phys. Rev. D} {\bf 23} (1981) 347.

\bibitem{Lyth}
A. R. Liddle and D. H. Lyth, in \textit{Cosmological Inflation and
Large-scale Structure} (C. U. P., Cambridge, 2000)

\bibitem{Lemoine}
M. Lemoine et al. , in \textit{Inflationary Cosmology} (Springer,
Heidelberg, 2008)

\bibitem{Komatsu}
E. Komatsu et al. , \textit{Astrophys. J. Suppl.} {\bf 180} (2009)
330

\bibitem{Nakayama}
Y. Nakayama, \textit{Int. J. Mod. Phys. A} {\bf 19} (2004) 2771


\bibitem{Vega}
C. Destri, H. J. de Vega and N. G. Sanchez, \textit{Phys. Rev. D}
{\bf 77} (2008) 043509

\bibitem{Albrecht}
A. Albrecht and R. H. Brandenberger, \textit{Phys. Rev. D} {\bf
31} (1985) 1225

\bibitem{Vilenkin}
A. Vilenkin, \textit{Phys. Rev. Lett.} {\bf 72} (1994) 3137

\bibitem{Riotto}
D. H. Lyth and A. Riotto, \textit{Phys. Rept.} {\bf 314} (1999) 1


\bibitem{Linde}
R. Kallosh and A. Linde, \textit{JCAP} {\bf 0704} (2007) 017

\bibitem{Boyano}
D. Boyanovsky, C. Destri, H. J. de Vega and N. G. Sanchez,
\textit{Int. J. Mod. Phys. A} {\bf 24} (2009) 3669

\bibitem{Rehman}
M. U. Rehman and Q. Shafi, \textit{Phys. Rev. D} {\bf 81} (2010)
123525

\bibitem{Shapo}
F. Bezrukov, A. Magnin, M. Shaposhnikov and S. Sibiryakov,
\textit{JHEP} {\bf 1101} (2011) 016

\end{thebibliography}
\end{document}